\RequirePackage{color}
\documentclass{JHEP3}
\usepackage{graphicx,amsmath,amssymb}
\usepackage{epsfig,multicol}
\usepackage{epsf}
\usepackage{epstopdf}
\input{epsf.sty}

\usepackage{graphicx}
\usepackage{amssymb}
\usepackage{amsmath}
\usepackage{float}

\def\half{{1\over 2}}
\def\({\left(}
\def\){\right)}
\def\[{\left[}
\def\]{\right]}

\def\e{\begin{equation}}
\def\q{\end{equation}}
\def\m{\begin{eqnarray}}
\def\n{\end{eqnarray}}

\title{Constraints on single-field inflation with \textit{WMAP}, SPT and ACT data--- A last-minute stand before \textit{Planck}}
\author{Cheng Cheng$^{1,2}$ \footnote{chcheng@itp.ac.cn}, Qing-Guo Huang$^1$ \footnote{huangqg@itp.ac.cn}, Yin-Zhe Ma$^{3,4}$ \footnote{mayinzhe@phas.ubc.ca}
\\\small{$^1$ \em
State Key Laboratory of Theoretical Physics, Institute of Theoretical Physics, Chinese Academy of Sciences, Beijing 100190, China}
\\\small{$^2$ \em
University of the Chinese Academy of Sciences, Beijing 100190, China}
\\\small{$^3$ \em Department of Physics and Astronomy, University of British Columbia, Vancouver, V6T 1Z1, BC Canada}
\\\small{$^4$ \em Canadian Institute for Theoretical Astrophysics, 60 St. George Street
Toronto, M5S 3H8, Ontario, Canada}
 }

\abstract{We constrain models of single field inflation with the
pre-\textit{Planck} CMB data. The data used here is the 9-year
Wilkinson Microwave Anisotropy Probe (\textit{WMAP}) data, South
Pole Telescope (SPT) data and Atacama Cosmology Telescope (ACT)
data. By adding in running of spectral index parameter, we find
that the $\chi^2$ is improved by a factor of $\Delta \chi^2=8.44$,
which strongly indicates the preference of this parameter from
current data. In addition, we find that the running of spectral
index $\alpha_s$ does not change very much even if we switch to
different pivot scales, which suggests that the power law
expansion of power spectrum is accurate enough till the 1st order
term. Furthermore, we find that the joint constraints on $r-n_{s}$
give very tight constraints on single-field inflation models, and
the models with power law potential $\phi^{p}$ can only survive if
$0.9 \lesssim p \lesssim 2.1$, so a large class of inflation
models have already been ruled out before \textit{Planck} data.
Finally, we use the $f_{NL}$ data to constrain the non-trivial
sound speed $c_s$. We find that the current constraint is
dominated by the power spectrum constraints which have some
inconsistency with the constraints from $f_{NL}$. This poses
important questions of consistency between power spectrum and
bispectrum of \textit{WMAP} data.}

\keywords{CMB, inflation}

\begin{document}

\section{Introduction}
The inflationary model \cite{Guth81,Linde82,Albrecht82} has
achieved a great success in modern cosmology, and it has been
confirmed by many high precision CMB and Large scale structure
experiments \cite{WMAP9,SPT,ACT}. It provides a good explanation
to a series problems such as flatness problem, horizon problem,
and monopole problem in the standard cosmology scenario. In
addition, inflation paradigm provides a natural explanation for
the origin of primordial perturbations which constitute the seeds
for the large scale structure we can see today. Therefore,
identifying the realistic inflation model becomes an important
task in observational cosmology.

Astronomical observations provide a large mount of data to
constrain the cosmological parameters, especially inflation
models. The default cosmology model people always use is the
``six-parameter'' $\Lambda$CDM cosmology model, in which the
canonical single-field slow-roll inflation (sound speed $c_s=1$)
is assumed in the model. However, the class of slow-roll inflation
models already have some weak tension with the observational data.
In Ref. \cite{WMAP9}, it is shown that the generic $\phi^p$
inflation model cannot provide consistent $r-n_{s}$ values within
reasonable range of number of \textit{e}-folds. In addition,
\textit{WMAP} 9-year data \cite{f_nl} suggests that the local
non-Gaussianity has a large positive value, while the orthogonal
non-Gaussianity is a large negative value, and these values are
hardly to be produced in the canonical single-field slow-roll
inflation models. Given these interesting tension between the
canonical single-field slow-roll inflation model and the current
observational data, we would like to explore the possibilities of
non-trivial sound speed $c_s \neq 1$ as well as non-zero running
of spectral index $dn_s/d \ln k$ to test their consistency with
current combination of \textit{WMAP} 9-year data \cite{WMAP9}, ACT
data \cite{ACT} and SPT data \cite{SPT}. We intend to finish this
work right before \textit{Planck} data release (expected on 21st
March, 2013) in order to make an immediate comparison before and
after the \textit{Planck} data. We hope that our work will
motivate theorists to explore more phenomena in the general
single-field slow-roll inflation model given the tight constraints
on single field inflation models.

\begin{figure}[H]
\begin{center}
\centerline{\includegraphics[bb=14 14 891 531,
width=12cm]{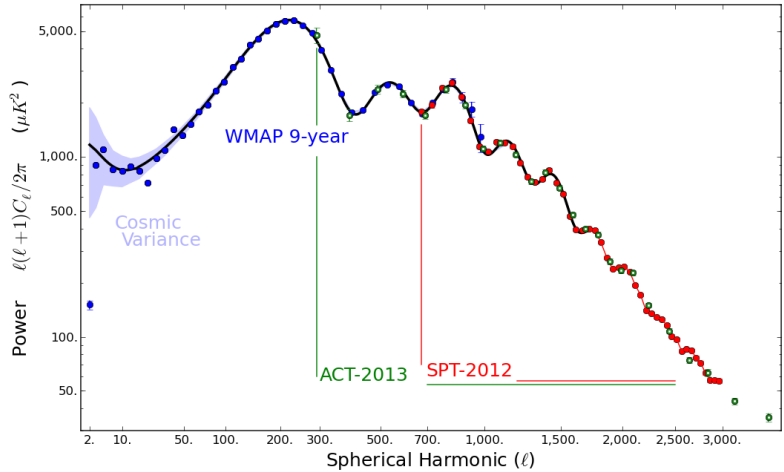}}
\end{center}
\caption{\textit{WMAP}9 temperature data with lensed ACT and SPT
data. \textit{WMAP}9, ACT and SPT data are mainly in the range
$2\leq l \lesssim 1000$, $300\lesssim l \lesssim 3000$, and $700
\lesssim l \lesssim 3000$ respectively. The theoretical curve is
the lensed CMB power spectrum with \textit{WMAP} 9-year
cosmological parameters and the light blue band is the cosmic
variance. The \textit{Planck} data will further tighten up the
error-bars in the middle regime. This figure is reprinted with
permission from Mark Halpern.}\label{fig:wmap9-act-spt}
\end{figure}

This paper is organized as follows. In Sec.~\ref{sec:methodology},
we will discuss the model we are focusing on, and the data we will
use to constrain the models. In Sec.~\ref{sec:results}, we will
present our results of fitting. The concluding remarks will be
presented in the last section.

% In this paper, we use the newest $f_{NL}$ data
%provided by WMAP9yr \cite{f_nl}, and the Cosmic Microwave
%Backgroud (CMB) power spectrum provided by WMAP9yr\cite{WMAP9},
%SPT\cite{SPT} and ACT\cite{ACT}, to investigate the constraint of
%comological paramters further. In Sec.2, we introduce the CMB data
%used and the methodology in the paper and in Sec.3, the fitting
%results will be represented. And we obtain the conclusion in
%Sec.4.

\section{Methodology}
\label{sec:methodology}

\subsection{The Model}
\label{sec:model}

We will use standard $6-$parameter $\Lambda$CDM model as our basic
model\footnote{The free running parameters are \{$\Omega_bh^2$,
$\Omega_ch^2$, $\Omega_{\Lambda}$, $\tau$, $n_s$, $A_s$\}, which
are fractional baryon, cold dark matter, and dark energy density,
optical depth, spectral index and amplitude of primordial scalar
perturbation respectively.}. We then allow $r$ (tensor-to-scalar
ratio), $dn_s/d\ln k$ (running of spectral index), $c_s$ (sound
speed for the curvature perturbation modes) to be varied since we
want to explore the level of constraints from these parameters.
The sound speed is related to the tilt of tensor power spectrum in
the general single-field inflation model through
\cite{Garriga:1999vw,Vazquez13} \footnote{Here we
assume ${c_s}$ as a constant. The perturbation mode with sound speed $c_s$ crosses horizon during inflation when $c_sk=aH$. Considering $n_s-1=-2\epsilon-\eta-s\sim {\cal O}(10^{-2})$ and both slow-roll parameters $\epsilon$ and $\eta$ are far less than 1, we conclude that $s\sim {\cal O}(10^{-2})$, where $s\equiv {\dot c_s\over H c_s}$. On the other hand, since $d\ln k\simeq Hdt$, $d \ln c_s/d\ln k=s$ and then  
$c_s(k)=c_s(k_0)\left(\frac{k}{k_{0}}\right)^s$ which shows that $c_s$ is roughly scale independent.}
\begin{equation}
n_{t}=-\frac{r}{8 c_{s}}, \label{eq:tensortilt} 
\end{equation}
where the tensor power spectrum is parameterized as
\begin{equation}
P_t(k)=A_{t}(k_{0}) \left(\frac{k}{k_{0}}\right)^{n_{t}},
\label{eq:tensorpower}
\end{equation}
Here $k_0$ is the pivot scale.
Thus the tensor to scalar ratio is defined as
\begin{equation}
r=\frac{A_{t}(k_{0})}{A_{s}(k_{0})}. \label{eq:r}
\end{equation}
Since tensor power spectrum also contribute to CMB angular power
spectrum $C^{TT}_{l}$ on very large scales, we will use the CMB
temperature angular power spectrum to constrain $r$ and $c_{s}$. For more discussion on how the sound speed $c_s$ changes the data fitting is given in \cite{previous}. 

In addition, we add the ``running of running'' parameter which
characterizes the running of running of spectral index, i.e.
\begin{equation}
\beta_s=\frac{d \alpha_s}{d \ln k}=\frac{d^{2} n_{s}}{d \ln k^2}.
\label{eq:running2}
\end{equation}
Thus the scalar power spectrum is parameterized as
\begin{equation}
P_s(k) =
A_s(k_{0})\left(\frac{k}{k_0}\right)^{n_s(k_{0})-1+\frac{1}{2}\alpha_s(k_{0})
\ln\left(\frac{k}{k_0}\right)+\frac{1}{6}\beta_s
\ln^2\left(\frac{k}{k_0}\right)}. \label{eq:power}
\end{equation}
Note that once the ``running of running'' ($\beta_s$) is
introduced into the model, the running of spectral index
$\alpha_s$ becomes a scale-dependent quantity. To remove any
ambiguity, we need to specify the pivot scale in the power law
expansion (Eq.~(\ref{eq:power})), this is why the $\alpha_s$ is
related to $k_{0}$. However, if $\alpha_{s}$ turns out to be less
dependent on $k_{0}$, it means that the truncation till $\alpha_s$
is enough (1st order), and there is no need to introduce a higher
order truncation ($\beta_{s}$).

The reason we want to release $\beta_s$ as the running of running
parameter is that SPT data \cite{SPT} gives a detection of a
negative value of the running of spectral index $\alpha_s= d n_s/d
\ln k$ at $k_{0}=0.025 \textrm{Mpc}^{-1}$. So we would like to add
this parameter as a higher order effect to monitor any possible
``running of running''. Even though it has not been detected, it
is expected to be significantly constrained and is useful for the
reconstruction of canonical single-field slow-roll inflation
\cite{Huang:2006hr,Huang:2006yt}.

\subsection{The data}
\label{sec:data}

We will use the most precise class of CMB data up-to-date, which
is the combination of \textit{WMAP} 9-year data \cite{WMAP9}, SPT
data \cite{SPT} and ACT data \cite{ACT}. The temperature angular
power spectrum from three data sets is shown in
Fig.~\ref{fig:wmap9-act-spt}. The combined data is named as
``$CMB$ data'' in the following discussion. We set the maximum
$l$-range of scalar model to be $7000$ ($l^{\textrm{s}}_{max}=
7000$), and maximal tensor $l-$range to be $3000$
($l^{\textrm{t}}_{max} = 3000$) in the running of MCMC chains. In
addition, we add Baryon Acoustic Oscillation data \cite{BAO} as
well as $H_{0}$ prior from HST (Hubble-Space-Telescope) project
\cite{HST} into our data source. In order to explore the variation
of sound speed, we add constrained $f_{NL}$ data provided by
\textit{WMAP} 9-year bispectrum into our likelihood. The sound
speed $c_s$ is related to the equilateral and orthogonal type of
non-Gaussianity $f_{NL}$ through Eq.(57) in \cite{f_nl}. So
according to \cite{f_nl} we assign $f_{i} =
(f_{NL}^{eq},f_{NL}^{orth})$ as the data vector, which is
\begin{eqnarray}
%%f_{NL}^{loc} = 37.2\pm19.9\    (-3 < f_{NL}^{loc} < 77\ at\ 95\%CL)\\
f_{NL}^{eq} = 51\pm136\        (-221 < f_{NL}^{eq} < 323\ at\ 95\% \textrm{CL}), \\
f_{NL}^{orth} = -245\pm100\    (-445 < f_{NL}^{orth} <-45\ at\
95\% \textrm{CL}). 
\end{eqnarray}
Then we use the $\chi^2$ function (Eq.(58) in\cite{f_nl}) to
calculate the best-fit value of $c_s$ \footnote{The parameter A in
Eq.(57) in \cite{f_nl} is running as a free parameter.}, i.e.
\begin{equation}
\chi^2 = \mathop{\Sigma}_{ij}f_iF_{ij}f_j-2
\mathop{\Sigma}_{i}F_{ii}f_i\hat{f_i}+\mathop{\Sigma}_{ij}\hat{f_i}F_{ii}F_{ij}^{-1}F_{jj}\hat{f_j}
\end{equation}
where $F_{ij}$ is the lower right four elements of the Fisher
Matrix
\begin{equation}
F = \left(
\begin{array}{ccc}
25.25& 1.06& -2.39\\
1.06& 0.54& 0.20\\
-2.39& 2.20& 1.00\\
\end{array}
\right) \times 10^{-4},
\end{equation}
and $\hat{f_i} =(51,-245)$.

For the extended model of $\Lambda$CDM, we will release $r$ and
$\alpha_s$ in the CAMB code \cite{CAMB} and further modify the
code to incorporate running of running parameter ($\beta_s$). We
run CosmoMC \cite{COSMOMC,cs} to generate MCMC samples. We will
express our results in term of best-fit value of marginalized
likelihood, as well as $1\sigma$ and $2\sigma$ confidence level
(CL) ($68.3\%$ and $95.4\%$ CL).

\section{Results}
\label{sec:results}
\subsection{Canonical single-field slow-roll inflation model ($c_s=1$)}

\subsubsection{$\Lambda$CDM cosmology model}
We first fix $c_s=1$ and investigate the constraints on parameters
$r$ and $\alpha_s$. The data sets we use here are $CMB$ data,
$BAO$ and $H0$. Here we consider ``$6-$parameter model'', ``$6-$parameter+$r$ model'',
``$6-$parameter+$\alpha_s$'' model and
``$6-$parameter+$r$+$\alpha_s$'' model which are expressed
as ``$\Lambda$CDM'', ``$\Lambda$CDM+$r$'', ``$\Lambda$CDM+$\alpha_s$'' and
``$\Lambda$CDM+$r$+$\alpha_s$'' models respectively.

\begin{figure}[h]
\centering
\includegraphics[width=10cm] {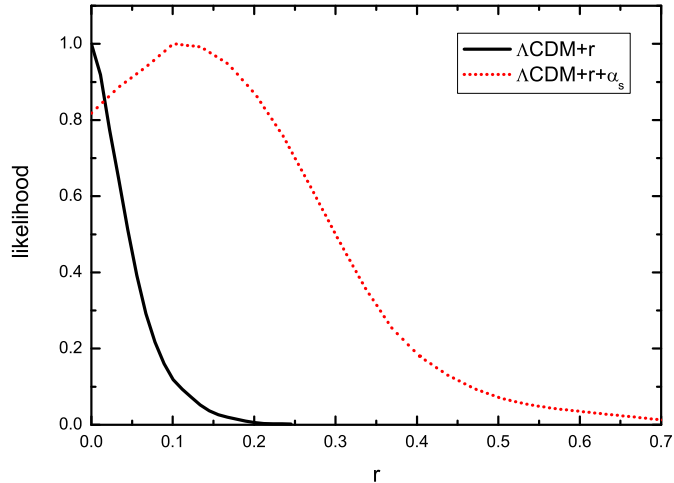}
\caption{\label{fig:compare_r} Likelihood of $r$ in case of
$\alpha_s$ fixed and $\alpha_s$ as free parameter. }
\end{figure}
In Fig.~\ref{fig:compare_r}, we can see that the likelihood of $r$
shifts a little if we switch $\alpha_s$ on and off. The solid line
is $\Lambda$CDM+$r$ model, and the dotted line is
$\Lambda$CDM+$r$+$\alpha_s$ model. In addition, the likelihood
becomes broader in $\Lambda$CDM+$r$+$\alpha_s$ model, and the
upper limit is also higher. This indicates that without the direct
polarization power spectrum, it is hard to draw concrete upper
limit on the amplitude of tensor mode $r$, since adding a single
extra-parameter can greatly broaden the constraint on $r$.

\begin{figure}[h]
\centering
\includegraphics[width=10cm] {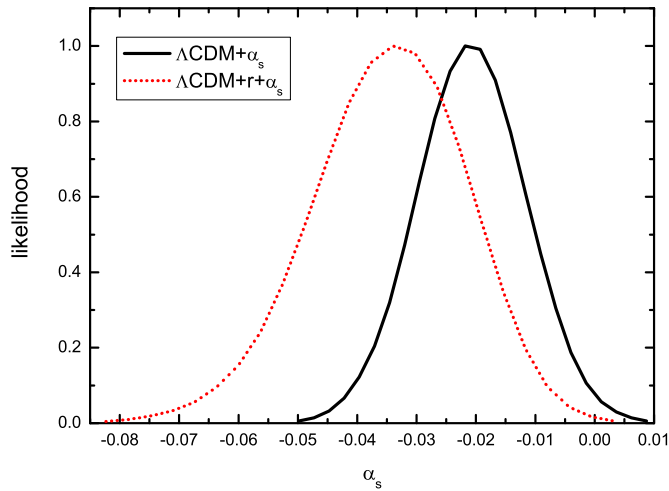}
\caption{\label{fig:compare_nrun} Likelihood of $\alpha_s$ in case
of $r$ fixed and $r$ free. }
\end{figure}
Similar thing exists in Fig.~\ref{fig:compare_nrun}. The solid
line is the $\Lambda$CDM+$\alpha_s$, and the dotted one is
$\Lambda$CDM+$\alpha_s$+$r$. One can see that the likelihood of
$\alpha_s$ is broader if $r$ is released as a free parameter. This
means that the two parameters have some level of degeneracy, which
is potentially able to be broken if the future polarization data
is added.

\begin{figure}[h]
\centering
\includegraphics[width=10cm] {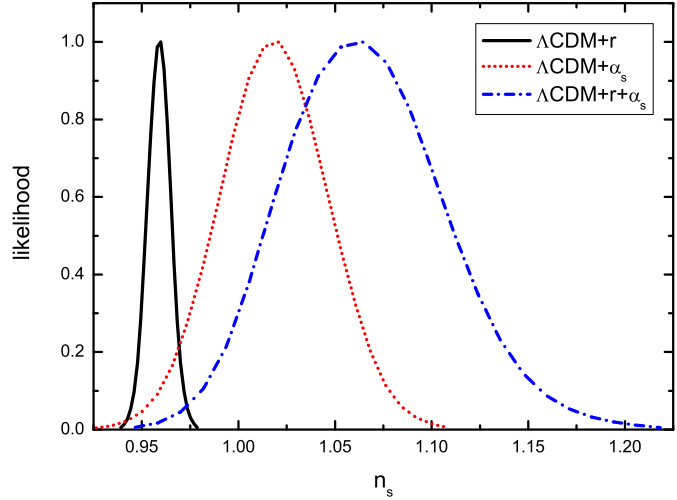}
\caption{\label{fig:compare_ns} Likelihood of $n_s$ in different
models. }
\end{figure}

The Fig.~\ref{fig:compare_ns} shows the likelihoods of $n_s$ for
three models. Here we consider all of the three models, i.e.
$\Lambda$CDM+$r$, $\Lambda$CDM+$\alpha_s$, $\Lambda$CDM+$r$+$\alpha_s$. We can see
that not only the peak of distribution shift, but also the range
of confidence level of $n_s$ changes quite a lot in three
different model: if we add $r$, the spectral index still prefers a
``red'' spectrum as $n_{s}<1$, but such situation does not exist
anymore in the case of $\Lambda$CDM+$\alpha_s$ and
$\Lambda$CDM+$\alpha_{s}$+$r$.

\begin{figure}[h]
\centering
\includegraphics[width=10cm,height=6cm] {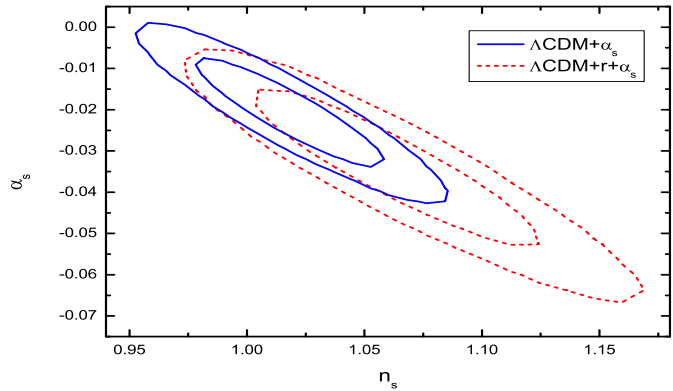}
\caption{\label{fig:ns-nrun} Joint constraint of $n_s$ and
$\alpha_s$ in $\Lambda$CDM+$\alpha_s$ (blue solid contours) and
$\Lambda$CDM+$\alpha_s$+$r$ model (red dashed contours). The
contours show $1\sigma$ and $2\sigma$ constraints. }
\end{figure}
Fig.~\ref{fig:ns-nrun} shows the contours of joint constraints on
$\alpha_s-n_s$ in $\Lambda$CDM+$\alpha_s$ model(the blue solid
curves) and $\Lambda$CDM+$r$+$\alpha_s$ model(the red dashed
curves). We can see adding $r$ leads to the shift of $n_s$ towards
bluer region, and the constraints become broader.

In Table~\ref{tab:cs=1}, we list the results of fitting by fixing
$c_s=1$. One can see that by introducing $\alpha_s$ parameter, the
$\chi^2$ really improve significantly $(\Delta \chi^2=4.22)$, indicating that the current
data prefer the inflation model with running of the spectral index.

%Comparing the results of $\Lambda$$CDM+r$, and
%$\Lambda$$CDM+r+\alpha_s$ model, which is shown in
%Fig.\ref{fig:compare_r}, Fig.\ref{fig:compare_ns} and
%Tab.\ref{tab:cs=1}, we can see when adding the parameter
%$\alpha_s$, credit ranges of both $r$ and $n_s$ is expanded, and
%the peaks of these two parameters shift a little, which means
%running $\alpha_s$ doesn't narrow the paramter space. The same
%conclusion can be got, when comparing the results of
%$\Lambda$$CDM+\alpha_s$ model and $\Lambda$$CDM+r+\alpha_s$ model
%to investgate the effect of parameter $r$, that adding $r$ can't
%constrain $\alpha_s$ and $n_s$ effectively.

%\begin{figure}[tbp]
%\begin{center}
%\includegraphics[width=8cm]{snplot.eps}
%\end{center}
%%\vspace{-0.8cm}
%\caption{The SN data plotted in Galactic coordinates.. The red
%points are moving away from us and the blue ones are moving
%towards us. The size of the points is proportional to the
%magnitude of the line-of-sight peculiar velocity. "X" is  our
%estimate of the direction of the tilted velocity estimated from
%the SN data.} \label{mock}
%\end{figure}

\begin{figure}[H]
\begin{center}
\centerline{\includegraphics[bb=0 0 626 405,
width=8.7cm]{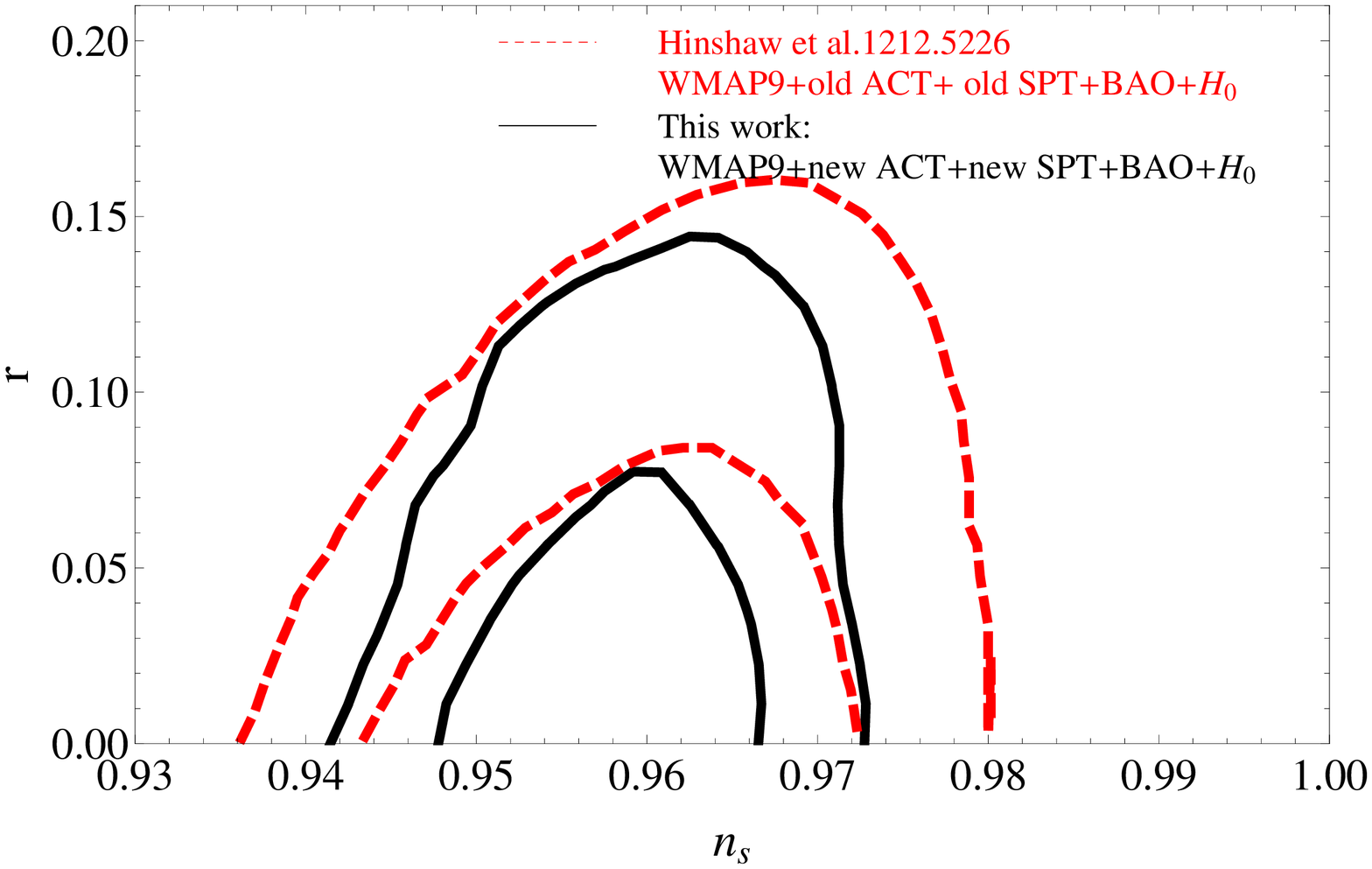}
\includegraphics[bb=20 25 284 215,
width=8.3cm]{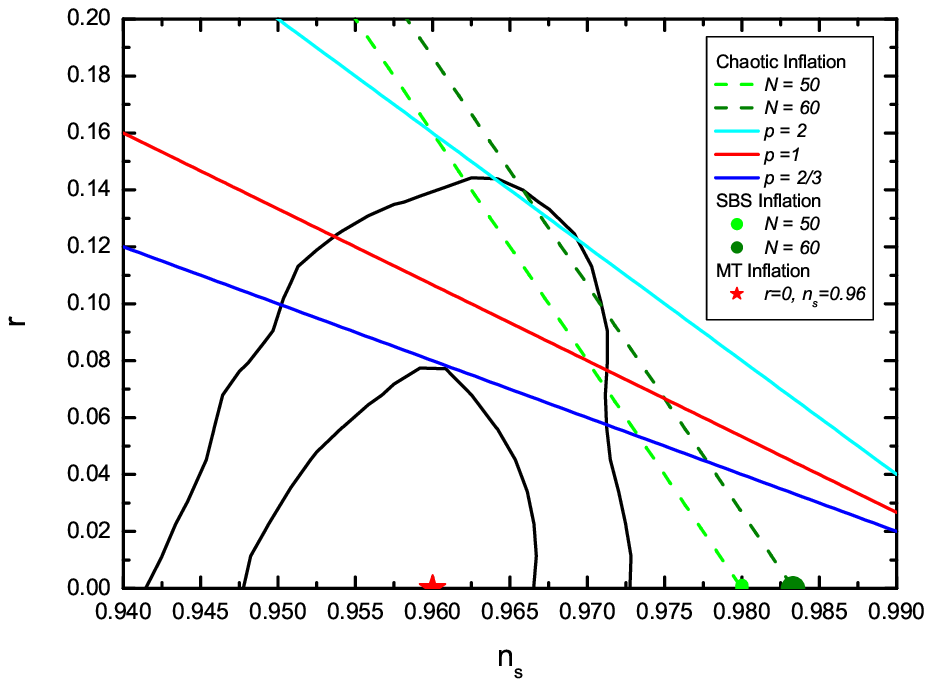}}
\end{center}
\caption{Joint constraint on $r-n_s$ and comparing with
\textit{WMAP}9+old SPT+old ACT (left panel) and model predictions
(right panel). Left panel: the red contours are from \textit{WMAP}
9-year results \cite{WMAP9} which combined \textit{WMAP}9+old
ACT+old SPT+BAO+$H_{0}$, while our constraints are the results of
\textit{WMAP}9+new ACT+new SPT+BAO+$H_{0}$. Right panel: we
consider several typical inflation models. (1) chaotic inflation
model \cite{Linde:1983gd} with potential $V(\phi)\propto \phi^p$.
The solid lines correspond to the predictions for different value
of $p$, and the shallow and darker green dashed lines correspond
to the predictions for $N=50$ and $N=60$ in the models with
different power index $p$. (2) spontaneously broken SUSY ({\bf
SBS}) inflation model whose potential is given by
$V(\phi)=V_0\(1+c\ln {\phi\over Q}\)$ which is assumed to be
dominated by $V_0$. (3) mass term ({\bf MT}) inflation model with
potential $V(\phi)=V_0-\half m^2 \phi^2$ where the mass term is
assumed to be subdominant.}\label{fig:r-ns}
\end{figure}

In the left panel of Fig.~\ref{fig:r-ns}, we compare our joint
constraints on $r$ and $n_{s}$ with the results from \textit{WMAP}
9-year paper \cite{WMAP9}. \textit{WMAP} 9-year results used
\textit{WMAP} 9-year data, combined with old SPT data, old ACT
data, BAO data and $H_{0}$ data and obtain the black contours
($1\sigma$ and $2\sigma$ CL). We used the similar combination,
except that our SPT and ACT data are the corresponding new data
sets \cite{SPT,ACT}. By updating the new data of ACT and SPT, one
can see that the constraints are tightened up to some extent. This
suggests that the new SPT and ACT data really provide a large
level arm for \textit{WMAP}9 data, which offer more constraining
power on small scales CMB angular power spectrum.

We use our results of joint constraints on plane of $r-n_{s}$ to
discuss its implication for inflation models (Right panel of
Fig.~\ref{fig:r-ns}).

%\begin{figure}[H]
%\centering \centerline{
%\includegraphics[bb=0 0 305 224, width=8.4cm]{nrun-run.eps}
%\includegraphics[bb=0 0 289 224, width=8cm]{run_like.eps}}
%\caption{Left: Joint constraints on $\alpha_s-\beta_s$. Right:
%Marginalized distribution of $\beta_s$ with $1\sigma$ CL $0.005
%\pm 0.021$.}\label{fig:nrun-run}
%\end{figure}

\begin{itemize}

\item Chaotic inflation model \cite{Linde:1983gd} whose potential
is given by $V(\phi)\propto \phi^p$. This model predicts
$r={4p\over N}$, $n_s=1-{p+2\over 2N}$, where $N$ is the number of
e-folds before the end of inflation. Given the current constraints
on the amplitude of inflation and the ``slow-roll'' parameter, $N$
is around $60$ but with some uncertainty of reheating process.
Here we take the range of $50$-$60$ as the reasonable range of
number of e-folds. The region between two dashed lines in
Fig.~\ref{fig:r-ns} indicates the prediction of chaotic inflation.
One can see that the models with $p=2$ \cite{Linde:1983gd} and
$p=2/3$ \cite{McAllister:2008hb} are disfavored at around
$2\sigma$ level, and only the models with $p\in [0.9, 1.8]$ for
$N=50$ or $p\in[1.5, 2.1]$
for $N=60$ are still consistent with data within $95\%$ CL. \\

\item
Spontaneously broken SUSY ({\bf SBS}) inflation model \cite{SUSY} with potential $V(\phi)=V_0\(1+c\ln {\phi\over Q}\)$, where the potential is assumed to be dominated by $V_0$ and $c\ll 1$. This model predicts $r=0$ and $n_s=1-{1\over N}$. The spectral index in this model is quite large and it is disfavored at more than $95\%$ CL. \\

\item Mass term ({\bf MT}) inflation model \cite{MT} with potential
$V(\phi)=V_0-\half m^2 \phi^2$ where the mass term is assumed to
be subdominant. The tensor-to-scalar ratio and spectral index in
this model are respectively given by $r=0$ and $n_s=1+2\eta$ where
$\eta=-m^2M_p^2/V_0$. This model can fit the data very well if
$\eta=-0.02$.

\end{itemize}

%In these models, $p$ is the slow-roll field power index, and $N$ is the e-foding number. The relation between $p$, $N$, $n_s$ and $r$ are $n-1=-(2+p)/2N$ and $r=4p/N$. The light green is for $N=60$, and the dark one is for $N=50$. The blue, red and cyan one is respective for $p=2$, $p=1$ and $p=2/3$. We can see the intersections of cyan with light green and with dark green are out of $n_s-r$ contour. It indicates that the model of $p=2/3$, which is derived from string theory, is disfavoured by current data.

\begin{table}
\centering
\renewcommand{\arraystretch}{1.2}
\scriptsize{ \caption{Results of fitting by fixing $c_s=1$. We set
$k_0 = 0.002 \rm{Mpc}^{-1}$, $l^{\textrm{s}}_{max}= 7000$, and
$l^{\textrm{t}}_{max} = 3000$ in the running of MCMC chains. }
\label{tab:cs=1} \
\begin{tabular}{cccccc}
\hline\hline
 &$\Lambda$CDM & $\Lambda$CDM+$r$ & $\Lambda$CDM+$\alpha_s$ & $\Lambda$CDM+$r$+$\alpha_s$ & $\Lambda$CDM+$r$+$\alpha_s+\beta_s$\\
\hline
$n_s$ &$0.961\pm0.007$ &$ 0.959\pm0.006$ & $1.018\pm0.027$ &$1.066\pm0.040$&$1.089\pm0.080$\\
$r(95\%CL)$&--&$<0.12$ & -- &$<0.42$&$<0.53$\\
$\alpha_s$ &-- &-- & $-0.021\pm0.009$&$-0.035\pm0.012$&$-0.050\pm0.057$\\
$\beta_s$ &--&--&--&--&$0.005\pm0.021$\\
Best fit -ln(Like)&$4921.52$& $4921.15$ & $4917.30$ & $4916.91$ & $4917.54$ \\
$\Delta{\chi^2}$ & $0$ & $-0.74$ & $-8.44$ & $-9.22$ & $-7.96$ \\
\hline
\end{tabular}
}
\end{table}

\subsubsection{Comparison of different pivot scale and the influence of running of runing of spectral index ($\beta_s$)}
\label{sec:alphas} 
\begin{figure}[H]
\centering
\includegraphics[width=10cm]{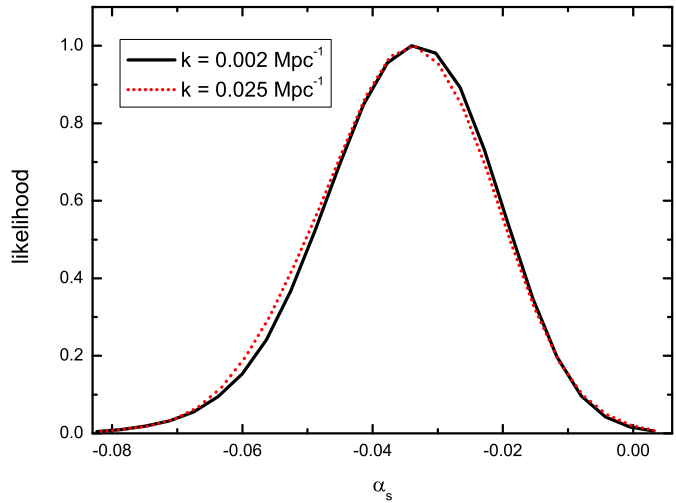}
\caption{The marginalized distribution of running of spectral
index $\alpha_s$ at different pivot scales.
\label{fig:nrun_diff_scale} }
\end{figure}

In the former sections, all the fittings are
done at pivot scale $k_0=0.002$ Mpc$^{-1}$ and the running of
spectral index is preferred at more than $2\sigma$ level. In this
section, we investigate the distributions of $\alpha_s$ at
different pivot scales. We use the model
$\Lambda$CDM+$r$+$\alpha_s$. The solid line is $k\ =\ 0.002\ {\rm
Mpc}^{-1}$, and the dotted line is $k\ =\ 0.025\ {\rm Mpc}^{-1}$
in Fig.~\ref{fig:nrun_diff_scale}. It shows that when the pivot
scale change, the distribution of $\alpha_s$ almost does not
change at all. This means that the constraints on $\alpha_s$ is
not sensitive to the pivot scale you choose, which indicates that
the truncation of power index expansion (Eq.~(\ref{eq:power})) is
accurate enough till 1st order.

\begin{figure}[H]
\centering \centerline{
\includegraphics[bb=0 0 305 224, width=8.4cm]{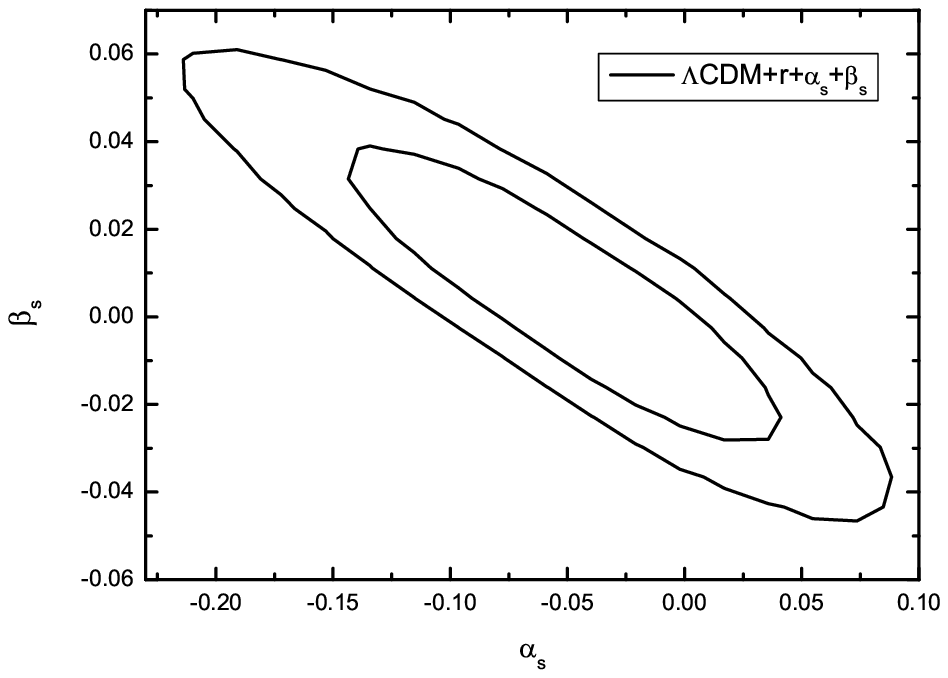}
\includegraphics[bb=0 0 289 224, width=8cm]{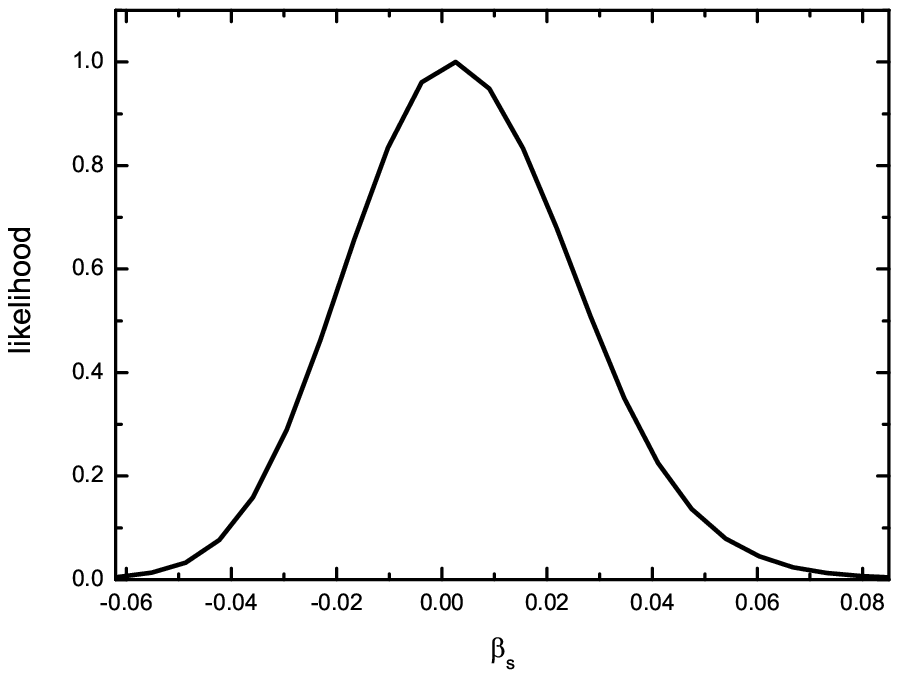}}
\caption{Left: Joint constraints on $\alpha_s-\beta_s$. Right:
Marginalized distribution of $\beta_s$ with $1\sigma$ CL $0.005
\pm 0.021$.}\label{fig:nrun-run}
\end{figure}

Considering the higher order power effect of the primordial
power spectrum, we introduce a new parameter $\beta_s$ to
characterize the ``running of running'' (Eqs.~(\ref{eq:running2})
and (\ref{eq:power})). Left panel of Fig.~\ref{fig:nrun-run} shows
the joint constraint on $\alpha_s$ and $\beta_s$, and the right
panel shows the marginalized distribution of $\beta_s$ with a flat
prior. We can see that the peak of $\beta_s$ slightly deviates
from $0$, but is perfectly consistent with zero within $1\sigma$
CL. This means that the current data do not support the ``running
of running of spectral index'', and therefore the power law
expansion of the scalar power spectrum (Eq.~(\ref{eq:power})) is
accurate enough till the $\alpha_s$ term. This is consistent with
what we find in Sec.~\ref{sec:alphas}. The fitting results are
shown in Table~\ref{tab:cs=1}.

%\end{figure}
%\begin{figure}[H]
%\centering
%\includegraphics[width=10cm]{run_like.eps}
%\caption{\label{fig:run_like} }
%\end{figure}

\subsection{General single-field inflation Model ($c_s$ free)}

\begin{figure}[H]
\centering
\includegraphics[width=10cm] {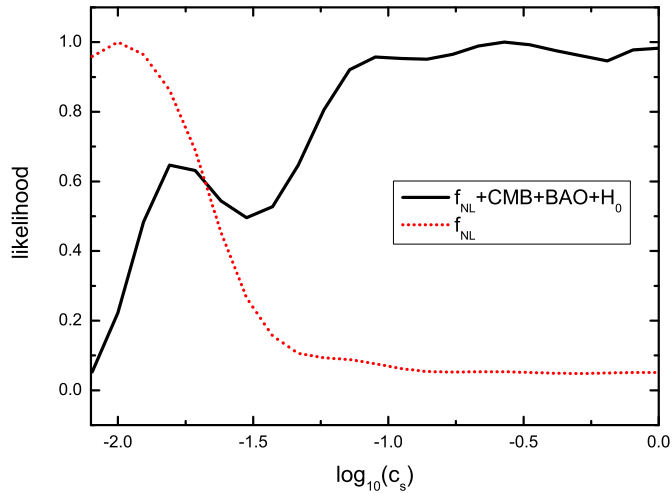}
\caption{\label{fig:cs_like} Likelihood of $c_s$ in different datasets.  }
\end{figure}

In this section, we release $c_s$ as a free parameter which is
constrained by the $f_{NL}$ data from \textit{WMAP} 9-year results
\cite{f_nl} \footnote{A similar constraint from \textit{WMAP}
7-year data and Background Imaging of Cosmic Extragalactic
Polarization (BICEP) experiment was given in \cite{previous}. }.
Table.~\ref{tab:B} shows the best fit (-log(Like)) and confidence
level of $n_s$, $r$ and $\alpha_s$ of two models. It can be seen
that adding $\alpha_s$ significantly reduces the best fit
-log(Like), but enlarge both the confidence interval of $r$ and
$n_s$.

Fig. \ref{fig:cs_like} shows the distribution of $c_s$. The dotted
line is the likelihood from $f_{NL}$ data while the solid line is
marginalized probabilities from the CMB+BAO+$H_{0}$+$f_{NL}$ data.
We can see that the $f_{NL}$ prefers a very low value of $c_{s}$
while the CMB data sets prefer a larger value, which indicates
some tension between each other. In addition, the combined
constraints are dominated by CMB power spectrum simply because the
number of CMB power spectrum data is far greater than the $f_{NL}$
data. The tension between the $f_{NL}$ data and the CMB power
spectrum data may
have a variety of indications: \\
1) it may indicate that the power
spectrum and bispectrum data are not consistent with each other,
which suggests that there are some uncleaned systematics in the
data sets; \\
2) it may also indicate that the underlying model,
i.e. single-field inflation cannot work at all when we confront it
with CMB power spectrum and bispectrum data. \\
In any case, we need to develop a method which can direct relate $c_{s}$ with power
spectrum and bispectrum (not just $f_{NL}$ data) and globally fit
this parameter by using full spectrum of CMB data. Such work is in
progress.

%they are not the same
%which should be the same in usual. It means the best-fit range of
%$c_s$ has low marginalized probability.\cite{cs} We infer that the
%effect on cosmological parameters of $c_s$ constraint by $f_{NL}$
%data is not as much as that of $CMB$ angular power spectrum. This
%is because there're only two points of $f_{NL}$, but more than two
%points of $CMB$ angular power spectrum, which leads to the
%unbalance. It indicates that there is a tension exist between
%$f_{NL}$ data and $TT$ angular power spectrum. We need to consider
%both the 3-point funcion and 2-point funcion to run $c_s$ as a
%free parameter, then constrain the $f_{NL}$ data.

\begin{table}[!htp]
\centering
\renewcommand{\arraystretch}{1.2}
\scriptsize{ \caption{Results of fitting with $c_s$ as a free
parameter. Here we use the same $k_{0}$ and $l_{max}$ as Table~1.}
\label{tab:B} \

\begin{tabular}{cccc}
\hline\hline
 & $\Lambda$CDM+$r$ & $\Lambda$CDM+$r$+$c_{s}$ & $\Lambda$CDM+$r$+$c_{s}$+$\alpha_s$\\
\hline
$n_s$ & $0.959\pm0.006$ & $ 0.958\pm0.006$ & $1.064\pm0.040$\\
$r(95\%CL)$& $<0.12$& $<0.11$ & $<0.40$\\
$\alpha_s$ & -- & -- & $-0.034\pm0.012$\\
Best fit -ln(Like) & $4921.15$ & $4922.14$ & $4918.17$\\
$\Delta{\chi^2}$ & $0$ & $1.98$ & $-5.96$\\
\hline
\end{tabular}
}
\end{table}

\section{Conclusion}
\label{sec:conclusion}

In this paper, we combine the most recent pre-\textit{Planck}
$CMB$ data to constrain the inflation model parameters. Our data
consists of \textit{WMAP} 9-year data \cite{WMAP9}, ACT data
\cite{ACT}, SPT data \cite{SPT}, Baryon Acoustic Oscillation data
\cite{BAO} as well as $H_{0}$ prior \cite{HST}. We mainly find
four interesting results from our numerical fitting:

\begin{itemize}

\item if we add in the running of spectral index $\alpha_{s}=d
n_{s}/d \ln k$, the $\chi^2$ value reduces a lot, which indicates
that it improves the fit to data very much.

\item By adding in a `3rd-order' parameter, i.e. running of
running $\beta_{s}$, we find that current data do not support
non-zero detection of $\beta_s$. In addition, by switching to
different pivot scales, the constraints on $\alpha_s$ do not vary
a lot. These two tests strongly suggest that the expansion of the
power spectrum is accurate enough till the 1st order ($\alpha_s$
term), and there is no observational hint for the higher order
scale-dependent terms.

\item Due to the new ACT and SPT data we used, our constraints on
$r-n_{s}$ is tighter than the \textit{WMAP} 9-year results
\cite{WMAP9}. Our constraints is already able to rule out a large
class of single-field inflation model even before \textit{Planck}
data. We show that the single field inflation with power law
$\phi^{p}$ can only survive if $p$ is in between $0.9$ and $2.1$,
and Spontaneously broken SUSY (SBS) inflation is ruled out firmly
by current observational data.

\item We release sound speed $c_{s}$ as a free parameter, and find
that the constraints on $c_{s}$ from $f_{NL}$ data and CMB power
spectrum are not consistent with each other. This strongly
indicates that either there is some unaccounted systematics in the
bispectrum data that may incur extra-error in the $f_{NL}$
estimation, or the model of varying $c_{s}$ cannot work at all
given these two datasets. In any case, this motivates us to
explore a set of formulism that directly compute power spectrum
and bispectrum given a $c_{s}$ value.

\end{itemize}

In conclusion, we find that pre-\textit{Planck} data have already
been able to set tight constraints on single field inflation
model. But current observational data still leave many open
questions to be solved. We hope such issues will be resolved when
the \textit{Planck} data becomes available in a few days.

%We find add inflation parameter will expand the credit range of
%inflation parameter. When considering the sound speed $c_s$, the
%umbalance between $CMB$ angular power spectrum and $f_{NL}$ leads
%to the $c_s$ tending to be $1$. It also indicates a tension
%between $CMB$ angular power spectrum and $f_{NL}$ data. It's
%better to combine the 3-point funcion and 2-point function to
%constrain $c_s$, then get $f_{NL}$ with $c_s$ constraint. We
%change the pivot scale, and find the peak of $\alpha_s$ a little
%shift. Then we add a new parameter to characterise this effect, we
%find it deviated from $0$, but more reliable result need more
%accurate data.

%section 1

\vspace{1.4cm}

%\noindent {\bf Acknowledgments}

%\vspace{.5cm}

\emph{Acknowledgement---} We would like to thank Mark Halpern to
share his figure with us. This work is supported by the project of
Knowledge Innovation Program of Chinese Academy of Science and a
grant from NSFC (grant NO. 10821504).

%\vspace{1.5cm}

%\appendix

%\section{appendix}
%\label{ap}


\newpage

\begin{thebibliography}{99}
%\baselineskip=16pt

%\vspace{0.7cm}


\bibitem{Guth81} A.~H.~Guth, Phys.\ Rev.\ D {\bf 23}, 347 (1981)
\bibitem{Linde82} A.~D.~Linde, Phys.\ Lett.\ B {\bf 108}, 389 (1982)
\bibitem{Albrecht82} A.~Albrecht and P.~J.~Steinhardt, Phys.\ Rev.\ Lett.\  {\bf 48},
1220 (1982).

\bibitem{WMAP9}
G.~Hinshaw, D.~Larson, E.~Komatsu, D.~N.~Spergel, C.~L.~Bennett, J.~Dunkley, M.~R.~Nolta and M.~Halpern {\it et al.},
%``Nine-Year Wilkinson Microwave Anisotropy Probe (WMAP) Observations: Cosmological Parameter Results,''
arXiv:1212.5226 [astro-ph.CO].
%%CITATION = ARXIV:1212.5226;%%

\bibitem{SPT}
K.~T.~Story, C.~L.~Reichardt, Z.~Hou, R.~Keisler, K.~A.~Aird, B.~A.~Benson, L.~E.~Bleem and J.~E.~Carlstrom {\it et al.},
%``A Measurement of the Cosmic Microwave Background Damping Tail from the 2500-square-degree SPT-SZ survey,''
arXiv:1210.7231 [astro-ph.CO].
%%CITATION = ARXIV:1210.7231;%%

\bibitem{ACT}
J.~L.~Sievers, R.~A.~Hlozek, M.~R.~Nolta, V.~Acquaviva, G.~E.~Addison, P.~A.~R.~Ade, P.~Aguirre and M.~Amiri {\it et al.},
%``The Atacama Cosmology Telescope: Cosmological parameters from three seasons of data,''
arXiv:1301.0824 [astro-ph.CO].
%%CITATION = ARXIV:1301.0824;%%

\bibitem{f_nl}
C.~L.~Bennett, D.~Larson, J.~L.~Weiland, N.~Jarosik, G.~Hinshaw, N.~Odegard, K.~M.~Smith and R.~S.~Hill {\it et al.},
%``Nine-Year Wilkinson Microwave Anisotropy Probe (WMAP) Observations: Final Maps and Results,''
arXiv:1212.5225 [astro-ph.CO].
%%CITATION = ARXIV:1212.5225;%%



\bibitem{Garriga:1999vw}
  J.~Garriga and V.~F.~Mukhanov,
  %``Perturbations in k-inflation,''
  Phys.\ Lett.\ B {\bf 458}, 219 (1999)
  [hep-th/9904176].

\bibitem{Vazquez13} J. A. Vazquez, M. Bridges, Y. Z. Ma, \& M. P.
Hobson, 1303.4014 [arXiv:astro-ph.CO]


\bibitem{previous}
C.~Cheng, Q.~-G.~Huang, X.~-D.~Li and Y.~-Z.~Ma,
  %``Cosmological Interpretations of Consistency Relation of Inflation Models with Current CMB Data,''
  Phys.\ Rev.\ D {\bf 86}, 123512 (2012)
  [arXiv:1207.6113 [astro-ph.CO]].


\bibitem{Huang:2006hr}
  Q.~-G.~Huang,
  %``Running of Running of the Spectral Index and WMAP Three-year data,''
  JCAP {\bf 0611}, 004 (2006)
  [astro-ph/0610389].

\bibitem{Huang:2006yt}
  Q.~-G.~Huang,
  %``Slow-roll reconstruction for running spectral index,''
  Phys.\ Rev.\ D {\bf 76}, 043505 (2007)
  [astro-ph/0610924].




\bibitem{BAO}
Beutler, F., {\it et al.} , Mon.\ Not.\ Roy.\ Astron.\ Soc.\ {\bf 416}, 3017 (2011);
Padmanabhan, N., Xu, X., Eisenstein, D. J., Scalzo, R., Cuesta,
A. J., Mehta, K. T., \& Kazin, E. 2012, [astro-ph.CO],arXiv:1202.0090;
L.Anderson, E.Aubourg, S.Bailey, D.Bizyaev, M.Blanton, A.~S.Bolton, J.Brinkmann and J.R.Brownstein {\it et al.}, Mon.\ Not.\ Roy.\ Astron.\ Soc.\  {\bf 428}, 1036 (2013)
Blake, C., {\it et al.} , Mon.\ Not.\ Roy.\ Astron.\ Soc.\ {\bf 425}, 405 (2012)

\bibitem{HST}
A.~G.~Riess , {\it et al.} , ApJ\ {\bf 730}, 119 (2011)

\bibitem{CAMB}
A.~Lewis, Phys. Rev. D {\bf 70}, 043011 (2004); A.~Lewis, A.~Challinor and A.~Lasenby, AJ {\bf 538}, 473 (2000); http://www.camb.info

\bibitem{COSMOMC}
A.~Lewis and S.~Bridle, Phys. Rev. D {\bf 66}, 103511 (2002);
http://cosmologist.info/cosmomc/


\bibitem{cs}
A.~Lewis and S.~Bridle,
  %{Cosmological parameters from CMB and other data: A Monte Carlo approach},
  Phys.\ Rev.\ D,{\bf 66},103511(2012)
  % doi = {10.1103/PhysRevD.66.103511},
  %url = {http://link.aps.org/doi/10.1103/PhysRevD.66.103511},
  %publisher = {American Physical Society}


\bibitem{Linde:1983gd}
  A.~D.~Linde,
  %``Chaotic Inflation,''
  Phys.\ Lett.\ B {\bf 129}, 177 (1983).

\bibitem{McAllister:2008hb}
  L.~McAllister, E.~Silverstein and A.~Westphal,
  %``Gravity Waves and Linear Inflation from Axion Monodromy,''
  Phys.\ Rev.\ D {\bf 82}, 046003 (2010)
  [arXiv:0808.0706 [hep-th]].

\bibitem{SUSY}
  G.~R.~Dvali, Q.~Shafi and R.~K.~Schaefer,
  %``Large scale structure and supersymmetric inflation without fine tuning,''
  Phys.\ Rev.\ Lett.\  {\bf 73}, 1886 (1994)
  [hep-ph/9406319];
  E.~J.~Copeland, A.~R.~Liddle, D.~H.~Lyth, E.~D.~Stewart and D.~Wands,
  %``False vacuum inflation with Einstein gravity,''
  Phys.\ Rev.\ D {\bf 49}, 6410 (1994);
  P.~Binetruy and G.~R.~Dvali,
  %``D term inflation,''
  Phys.\ Lett.\ B {\bf 388}, 241 (1996)
  [hep-ph/9606342];
  E.~D.~Stewart,
  %``Inflation, supergravity and superstrings,''
  Phys.\ Rev.\ D {\bf 51}, 6847 (1995)
  [hep-ph/9405389];
  D.~H.~Lyth and A.~Riotto,
  %``Particle physics models of inflation and the cosmological density perturbation,''
  Phys.\ Rept.\  {\bf 314}, 1 (1999)
  [hep-ph/9807278].

\bibitem{MT}
  E.~D.~Stewart,
  %``Flattening the inflaton's potential with quantum corrections,''
  Phys.\ Lett.\ B {\bf 391}, 34 (1997)
  [hep-ph/9606241];
  E.~D.~Stewart,
  %``Flattening the inflaton's potential with quantum corrections. 2.,''
  Phys.\ Rev.\ D {\bf 56}, 2019 (1997)
  [hep-ph/9703232].





\end{thebibliography}
\end{document}